\documentclass[12pt]{article}
\usepackage{amsmath,amsthm,amssymb}
\usepackage{fullpage}
\usepackage{setspace}

\theoremstyle{plain}
\theoremstyle{remark}

\newcommand{\x}{\mathbf{x}}
\newcommand{\y}{\mathbf{y}}
\newcommand{\thet}{\boldsymbol{\theta}}

\newcommand{\R}{\mathbb{R}}
\renewcommand{\S}{\mathcal{S}}

\title{The Relativity of Theory}
\author{Nisheeth Srivastava}

\begin{document}
\maketitle
\begin{abstract}
A general information-theoretic framework for deriving physical laws is presented and a principle of informational physics is enunciated within its context. Existing approaches intended to derive physical laws from information-theoretic first principles are unified as special cases of this framework with the introduction of constraints dependent on the physical process of observation. Some practical, theoretical and epistemological implications of the validity of this approach are examined.
\end{abstract}
\section{Introduction}
\label{sec:intro}
The consideration of information (subjectively, awareness) as a fundamental ontological entity is a view with an extremely illustrious pedigree in several Idealistic traditions of philosophy\footnote{In the Western traditions, see e.g., Neoplatonism, Kant's Transcendental Idealism and Berkeley's Immaterialism. \emph{Advaita} and \emph{Zen} are good examples from the East. Sufism and Gnosticism are good representatives from the Semitic traditions.}. In the past half century, this philosophical position has been adopted by some scientific/computational programs and specified to various degrees of rigor in various representational structures, e.g.  \cite{whee90,zuse70,wolf02}. While many of these observers have designed different perspectives on the subject, the theme of a fundamental informational nature of reality runs through all their narratives. These are, in general terms, the philosophical approaches to informational physics. They begin with an ontological assumption regarding the nature of reality and attempt to find representational theories of sufficient expressiveness and precision to be able to make testable statements in the physical world.

In a different epistemological context, beginning with Frieden \cite{fried90}, there have been several sporadic efforts (see, e.g., \cite{fried04, skal05, lisi06}) made to re-derive the foundational principles of quantum mechanics with a probabilistic and/or information-theoretic basis, making no physical assumptions. This effort is philosophically appealing because of the nebulous understanding of the role of the observer and the observer's effect on the wave function in the standard Copenhagen interpretation. Furthermore, since most theories of physics, including the Standard Model, are generally expressed as quantum field theories, an information-theoretic interpretation of QFT, if justifiable, directly affects the entire epistemological basis of physics. As a matter of fact, a controversial information-theoretic unification model has already been proposed \cite{fried04} by Frieden as an extension to his original approach \cite{fried90}. These constitute what we call the mathematical approaches to informational physics. They begin from an agnostic epistemic viewpoint, operating entirely within existing representational frameworks of probability and information and attempt to characterize the ontological assumptions necessary for their mathematical findings to be consistent with existing theories of reality in the same mathematical representation.

In this brief note, we examine prospects of unifying the philosophical and mathematical currents of informational physics. We do so by formulating the problem of information optimization in a general mathematical setting. We then discuss how the simple selection of an information measure (frequentist probability) and its physical interpretation allows us to recover statistical mechanics from our general framework, an insight first gleaned by Jaynes \cite{jayn63}. We remark on the prospect of employing a similar approach to deriving quantum mechanics for large systems, viz., quantum field theories, an observation independently made by Lisi \cite{lisi06}. We further note how a different choice of information measure (Fisher Information) recovers Frieden's derivation of Schrodinger's equation \cite{fried91} as well as (within some interpretative latitude) his broader Extreme Physical Information framework \cite{fried04} of informational physics. Finally, using heuristic arguments, we suggest that this apparent unity offers us insight into both the nature of physical laws as well as the surprising comprehensibility of physics using mathematics \cite{wign60}.

\section{Information optimization}
\label{sec:inf}
Let $\S$ represent the set of possible outcomes in an isolated system under observation, where every outcome $s\in\S$ is a set of variables that sufficiently represent the system behavior for each outcome. For example, for a classical point particle (position $x$, momentum $p$) moving in a straight line, $\{x,p\}$ is a sufficient representation. In order to study changes in outcomes, we consider each outcome $s$ to be a continuous function of a set of parameters $t$.

If we were to consider the problem of obtaining knowledge in the broadest epistemic sense, an observer would be ignorant of the membership of $\S$. To be able to make meaningful statements about information optimization, we require that an observer possess information about the existence of at least some $s\in\S$\footnote{This is equivalent to saying that no meaningful discussion of a coin-tossing experiment can commence unless the subject knows it has two outcomes. He need not know that the two outcomes occur with equal probability.}. The goal of the observer is two-fold. In the first place, he wishes to explain the relative frequency of occurrence of different outcomes of the system with a \emph{theory}. We consider a good theory to be one which explains the relative frequencies of the occurrence of outcomes with a set of statements of a much lower Kolmogorov complexity  \cite[Ch.1-2]{grun05} than the set of statements required to describe the occurrence of outcomes. In the second, he would like to be able to make predictions regarding the behavior of unobserved outcomes of the system based on his observation of past outcomes. Note that these two aspects of \emph{episteme} need not be related, e.g., the true theory regarding a fair coin,`This is a random process', implies that prediction becomes impossible, whereas predicting a coin will show up heads (or tails) because it has shown up heads thrice before is a process fraught with peril\footnote{As a quick visit to the nearest roulette table at one's friendly neighborhood casino would readily demonstrate!}.

For either (or both) of these purposes, the observer associates a level of certainty or uncertainty with each possible outcome of the system. Let us denote this with an information measure $f(s)$, the choice of which will determine the form of the solution to the information optimization problem. Furthermore, since multiple outcomes are feasible, the observer must construct an aggregate measure of certainty (or uncertainty). In general, we can denote this by $F:\S'\rightarrow\R$, where $\S'\subseteq\S$ and depends on the observer's knowledge of the membership of the system.  

Now that the observer has quantified the extent of his knowledge/ignorance of the system, he must adjust these quantities to account for observations of system outcomes that he makes. The method of adjustment employed, which is essentially \emph{inference} may be arbitrary, but a principled approach would be to employ \emph{Bayesian} inference. While this might appear to constrain the generality of our informational framework, it must be noted that the Bayesian formalism has been shown to be quite generally applicable and along with the axioms of probability, appears to be quite fundamental to the way in which humans naturally think \cite{tene01}. Therefore, to assume that the system of updates that the observer employs will be Bayesian in nature does not appear to be overly restrictive.

Note further that Bayesian inference has been shown to be equivalent \cite{cati06} to the principle of maximum relative entropy \cite{shor81}, an extension of Jaynes' famous original MaxEnt principle \cite{jayn63} applicable to uniform prior distributions to arbitrary priors. MaxEnt, therefore, is a special case of Bayesian inference, where one begins with an agnostic belief regarding one's current state of knowledge. This, naturally, exposes the argument to the same opposition that proponents of MaxEnt face in general. It should therefore be emphasized that the MaxEnt assumption is not a crucial aspect of this framework and is used here to simplify our analysis in the examples that we consider. The general Bayesian argument (or its maximum relative entropy equivalent) constitute our general system of updates.

Having defined our terms of reference, we can now state our informational physics hypothesis somewhat more formally as a principle assumed in all our subsequent derivations. The principle of informational physics states that \emph{an observer's task of developing a theory concerning the evolution of the system characterized by a space of outcomes $\S$ is equivalent to extremizing a measure of his information/ignorance $F$ constrained by what he \emph{believes-he-knows}\footnote{The problem of distinguishing between \emph{believe-he-knows} and \emph{knows} is one that we will not address in this paper.} about the system. Known physical laws emerge as mathematically optimal solutions to this information extremization procedure.}

\section{Physical laws}
\subsection{Realistic information measures}
\label{sec:meas}
While a plethora of information/uncertainty measures (see e.g. \cite{tops08}) can be used within the information framework defined in Section \ref{sec:inf}, the one that is most commonly used in learning theory is Shannon entropy with an underlying frequentist definition of probability, i.e., assuming $p[s]$ measures the frequency of occurrence of outcome $s$ in the space of known outcomes $s\in\S'$, the observer's uncertainty $f(s)$ with respect to the outcome $s$ is given by $- p[s]\log p[s]$ and the aggregate uncertainty across all known paths $F(\S')$ as
\begin{equation}
F(\S') = -\sum_{s\in\S'}{p[s]\log p[s]} = -\int_{s\in\S'}{p[s]\log p[s]~ds}.
\label{eqn:entropy}
\end{equation}

The $F$ obtained here is precisely the Shannon entropy of the system and is a measure of the uncertainty of the observer regarding the behavior of the system. Concomitantly, it is possible to define $F(s)$ as the information divergence of the observer's hypothesis concerning the system to the best\footnote{This notion can be formalized in terms of compactness, predictability, some combination thereof or in some other way.} theory. Representing both the observer's hypothesis and the target theory as distributions over the space of outcomes, we can use the standard definition of information divergence (alternatively Kullback-Leibler divergence)  to obtain an aggregate information measure for the system, viz.,
\begin{equation}
F_{b}(\S') = \int_{s\in\S'}{a[s]\log\frac{a[s]}{b[s]}~ds},
\label{eqn:dive}
\end{equation}
where $b[s]$ is the observer's hypothesis and $a[s]$ is the target theory.

Note that in this case, minimizing the information measure would result in the observer finding a theory close to the optimal. Parenthetically, this recovers the MDI principle of model selection first proposed by Kullback and subsequently extended via the AIC criterion \cite{akai74} to statistical model selection \cite{burn02}.

The third important information information measure we consider here is a variant of Fisher Information first proposed in \cite{fried90}]. Frieden \cite{fried04} demonstrates that assuming $n$ independent measurements of outcomes as well as shift invariance in the system, it is possible to construct an information measure of the form
\begin{equation}
F(s) = 4\int_{s\in\S'}{\sum_{n}{\frac{\partial q}{\partial s}\cdot\frac{\partial q}{\partial s}}~ds},
\label{eqn:fisher}
\end{equation}
where $p[s] = q^2[s]$.

As we mention above, these are by no means the only information measures available. A wide variety of distance (e.g. Levenshtein, Hellinger, Dudley) and divergence measures (e.g., the Bregman family of divergences) can be used to replace (\ref{eqn:dive}) \cite{tops08}. Much work has been done to generalize entropy measures e.g., Tsallis entropies, and multiple definitions of probability abound. Thus, there are several information measures that may be used in place of the ones we have considered\footnote{We feel, however, that such specific efforts may not be very useful in practice. A more general view of information, on the other hand, would lack the mathematical rigor of our construction.}. The optimization problem and its interpretation must necessarily remain the same. There is thus, much flexibility in this framework to allow for physical theories to be constructed using different representations of information.

\subsection{Statistical and quantum mechanics}
\label{sec:shannon}
From the arguments presented in Section \ref{sec:inf}, it is evident that the task of constructing a physical theory for observations made on a physical system is equivalent to solving an inference problem given constraints on the observation process. Working with the entropic formulation of $F$ as defined in (\ref{eqn:entropy}), we now define minimal constraints on the inference problem as observed in the process of physical measurements.

For physical systems, it appears appropriate to conjecture that the probabilities $p[s]$ of all known outcomes should sum to one, .i.e., the observer should believe that he knows $\S$ (when, in fact, he only knows $\S'$). Thus, the observer \emph{believes-he-knows}
\begin{equation}
\sum_{s\in\S'}p[s(t)] = \int_{s\in\S'}{p[s]~ds} = 1.
\label{eqn:exist}
\end{equation}
Furthermore, in order to develop his theory, the observer, given multiple observations, \emph{believes-he-knows} how the system has been behaving. Mathematically, this means that he \emph{believes-he-knows} the expected value $\langle A\rangle$ of the \emph{history} $A[s]$ of the physical system given multiple observations, where the \emph{history} may be defined in any mathematical manner that takes the past behavior of the system corresponding to outcome $s$ into account. That is,
\begin{equation}
\bar{A} = \langle A\rangle = \int_{s\in\S'}{A[s]p[s]~ds}.
\label{eqn:measure}
\end{equation}

It is instructive to note here that the choice of information measure is independent of the physical constraints imposed, and that the physical constraints imposed are consequent to the process of observation, not the physical system being measured.

Recovering the optimal theory for the system using (\ref{eqn:entropy}) as an information measure and (\ref{eqn:exist}), (\ref{eqn:measure}) as informational constraints is now equivalent to extremizing,
\begin{align*}
F' &= - \int_{s\in\S'}{p[s]\log p[s]~ds} + \lambda\left(1 - \int_{s\in\S'}{p[s]~ds}\right) + \nu\left(\bar{A} - \int_{s\in\S'}{p[s]A[s]~ds}\right),\\
&= \lambda + \nu\bar{A} - \int_{s\in\S'}{(p[s]\log p[s] + \lambda p[s] + \nu p[s]A[s])~ds}.
\end{align*}

Differentiating with respect to $p[s]$ and setting the derivate to zero gives us$$ \int_{s\in\S'}{(\log p[s] + 1 + \lambda + \nu \bar{A}[s])~ds} = 0,$$
which gives us an expression for $p[s]$ of the form
\begin{equation}
p[s] = \frac{1}{Z}~e^{-\nu \bar{A}[s]}.
\label{eqn:solve}
\end{equation}
Differentiating with respect to $\lambda$, $Z = 1/e^{-1 - \lambda}$ which gives us,
\begin{equation}
Z = \int_{s\in\S'}{e^{-\nu \bar{A}[s]}~ds}.
\label{eqn:partition}
\end{equation}
Similarly, we can determine $\nu$ by solving the equation
\begin{equation}
-\frac{\partial}{\partial\nu}~\log Z = \bar{A}.
\label{eqn:constt}
\end{equation}

Now, note that Z, as defined in (\ref{eqn:partition}) has the structure of a partition function, while $\nu$ plays the role of an intensive variable (does not depend on the cardinality of $\S'$). The physical interpretation of this partition function may be inferred from the context of the domain of the physical system. In the case of classical statistical ensembles, we specify the \emph{history} of the system in a configuration $s$ to simply be the energy in that particular state and take the intensive variable to be the familiar $k_B T$. By doing so, we recover Jaynes' information-theoretic formulation of thermodynamics in the form of the well-known statistical canonical ensemble
$$Z = \sum_{states}{e^{-\frac{1}{k_B T}E[state]}}.$$

As Lisi points out, the formal similarity of the quantum partition function to the statistical canonical ensemble implies that it can be derived along similar lines. Following Lisi \cite{lisi06}, we interpret the notion of \emph{history}, in the case of a quantum ensemble, as the action $S$ corresponding to a particular \emph{path} as understood in Feynman's path integral formulation. The intensive variable is taken as $\imath\hbar$. By doing so, we recover the quantum partition function\footnote{Since it is possible to subsume all possible quantum actions in the informational framework we have discussed in Section \ref{sec:inf}, it is tempting to declare victory for informational physics here. However, since we do not have a characterization of the class of actions that have physical interpretations, we are far from being in a position to do so.},
$$Z = \sum_{paths}{e^{-\frac{1}{\imath\hbar}S[path]}}.$$
Thus, we see that using an entropic formulation of uncertainty about the possible outcomes of a physical system allows us to derive partition functions that, afforded reasonable physical interpretations, allow us to recover existing physical laws and explain the similarity in their formal structure with a deeper informational hypothesis.

\subsection {Extreme Physical Information}

In Section \ref{sec:shannon}, we have seen that using Shannon entropy as a measure of uncertainty allows us to recover existing physical laws for macroscopic ensembles. Beginning from the observation that a large number of physical laws are phrased in the form of second order differential equations, Frieden \cite{fried04} has suggested the possibility of deriving these laws through a general framework called Extreme Physical Information (EPI) which operates on a principle of Fisher Information extremization. In this Section, we examine the use of a Fisher Information based measure in our general informational framework and find that our approach replicates the basic mathematical structure of EPI without having to make any of its potentially questionable metaphysical assumptions\footnote{In particular, we remove the necessity of an observer-Nature information game, the concept of `bound' information and the necessity of Fourier transforms as a descriptor of symmetry}.

Recall that the standard definition of Fisher Information is of the form
\begin{equation}
I = \int{\left(\frac{\partial\log p(y|\thet)}{\partial\thet}\right)^2 p(y|\thet)~dy},
\end{equation}
where $p(y|\thet)$ is some interpretation (we assume frequentist) of the probability of $y$ conditioned on $\thet$. Fisher Information measures the amount of information that the random variable $y$ carries about the unknown parameter $\thet$. In the context of physical systems, we take $y$ to represent the value of physical measurements and $\theta$ to be the \emph{true}\footnote{In the informational context, `true' is equivalent to the value predicted by the best theory.} value of the physical quantity being measured. Geometrically speaking, the informativeness in this context is measured as a function of the steepness of the probability density function near the maximum likelihood estimate of the unknown parameter. Thus, unlike Shannon entropy, using Fisher Information as an uncertainty measure allows us to take local structure in the definition of the support of the space of outcomes $\S$ into account. Finally, note that since Fisher Information, like Kullback-Leibler divergence, measures informativeness in comparison to an ideal theory, finding a good physical theory in this estimation framework is equivalent to minimizing Fisher Information over the relevant domain. Without having to resort to an observer-Nature information game (observer maximizes information, Nature maximizes error), as posited in, e.g., \cite{fried04,tops08}, we recover the basic mathematical principle of EPI.

In general, the unknown parameter $\thet$ as well as the measurement $\y$ may be a $V$-dimensional vector, e.g., the position of a point particle in $\mathbb{R}^V$. In that case we obtain a $V\times V$ Fisher Information matrix. In physical measurements, the dimensions of the unknown parameter will often be orthogonal in the sense that the MLEs will be independent. Thus, the off-diagonal elements of the Fisher Information matrix will be zero, and the trace will be a comprehensive measure of the informativeness of the estimation. It may also be necessary to account for the effect of multiple observations $\y^{(n)}: n = 1\cdots N$ on the informativeness of the estimation procedure for the unknown parameter $\thet_v^{(n)}$ at the $n^{th}$ instance. Following Frieden, the sum of the traces of the $V\times V$ matrix corresponding to each observation can be shown to be an upper bound on the Stam information of the system and is thus a bound on the capacity of the estimation procedure to convey information about the measured quantity. Also, from the assumption of independence, the joint probability $p(\y|\thet)$ decomposes into a product of marginals $\prod_{n=1}^{N}{p_n(\y^{(n)}|\thet^{(n)})}$, resulting in the following simplification,
\begin{align*}
F &= \sum_{n=1}^{N}\int{p(y|\theta)\sum_{v=1}^{V}\left(\frac{\partial\log p(\y|\thet)}{\partial\thet_{v}^{(n)}}\right)^2~d\y^{(n)}},\\
&= \sum_{n=1}^{N}\int{\frac{1}{p_n}\sum_{v=1}^{V}\left(\frac{\partial p_n}{\partial\thet_v^{(n)}}\right)^2~d\y^{(n)}.}
\end{align*}
Finally, we assume shift invariance for the measurement process, i.e., the error in measurement $\x^{(n)} = \y^{(n)} - \theta^{(n)}$ is independent of the value of $\thet$. Then, $p_n(\y^{(n)}|\thet^{(n)}) = p_n(\x_n)$. To recapitulate, we have shown that it is possible to define a scalar information measure based on Fisher information that depends entirely on the fluctuations in the measurement process. By replacing the probability density $p_n$ with the `real' probability amplitude $q_n$ such that $p_n = q_n^2$ and suppressing the index $v$ in the notation (assuming measurements are made with no preferred dimension), we get
\begin{equation}
F = 4\int{\sum_{n}\nabla q_n\cdot\nabla q_n~d\x_n},
\label{eqn:frieden}
\end{equation}
which is equivalent to (\ref{eqn:fisher}) with some additional physical interpretation of the outcome space $\S$ attached. Specifically, whereas (\ref{eqn:fisher}) measures informativeness in a general space of outcomes $s\in\S$, (\ref{eqn:frieden}) measures informativeness corresponding to a physical measurement process where the form and the physical interpretation of the outcome $\x = \y - \thet$ is well-specified. Since physical laws dealing with idealized measurements are largely concerned with this specific class of outcomes, we use (\ref{eqn:frieden}) in the remainder of this Section.

Using the principle of informational physics, deriving a physical law using (\ref{eqn:frieden}) is equivalent to minimizing the information measure $F$ under constraints on the process of measurement. We formulate these constraints following our treatment of the Shannon entropy case in Section \ref{sec:shannon}. Since the information measure $F$ has no outcome-specific probabilistic interpretation, we will not have a density function normalization constraint analogous to (\ref{eqn:exist}). Analogous to (\ref{eqn:measure}), however, we assume here that the observer, given multiple observations, \emph{believes-he-knows} how the system has been behaving. In this context, this translates into the assumption that the observer's empirical estimate of the expected \emph{history} of all possible outcomes of the system is accurate, i.e.,
\begin{equation}
\bar{A} = \langle A\rangle = \int{A(\x) p(\x) d\x}.
\end{equation}

Thus, the informational quantity to be minimized takes the form,
\begin{equation}
F' = 4\int{\sum_{n}\nabla q_n\cdot\nabla q_n~d\x_n} + \lambda\left(\bar{A} - \int{A(\x) p(\x) d\x}\right),
\label{eqn:epi}
\end{equation}
which is structurally equivalent to the $K = I - J$ Lagrangian of EPI \cite{fried04}. Comparing this derivation of the EPI Lagrangian with that described in \cite{fried90} in the specific context of the derivation of the time-independent Schrodinger's equation and subsequently explicated in \cite{fried04} to encompass a general framework that results in information-theoretic derivations of the Dirac, Klein-Gordon and Maxwell's equations, we see that our informational framework not only subsumes the mathematical structure of EPI, but also motivates it in a far more convincing manner by eliminating the need for defining a bound information functional \cite{fried90} as well as the requirement of taking a Fourier transform of the original information measure \cite{fried04} to do so. Thus, we find that we have removed a significant number of the \textit{ad hoc} procedural assumptions of EPI while recovering its mathematical structure as a special representational implementation of a more general informational physics framework.

As an example of the application of the EPI method to deriving physical laws, we can briefly interpret Frieden's derivation of Schrodinger's equation (in one-dimension) starting from (\ref{eqn:epi}). The history $\bar{A}$ is assigned a physical interpretation of kinetic energy of the particle (potential energy would appear as a uniform background energy to an observer concerned solely with measuring the position of the particle). Recall that Schrodinger's equation measures probability using complex probability amplitudes. These may be constructed from real probability amplitudes $q_n(x)$ $$\psi_{n} = \frac{1}{\sqrt{N}}(q_{2n-1} + \imath q_{2n}), n = 1\cdots N/2,$$ with no loss of generality. Since the kinetic energy over N measurements is being considered, we work with the quantity $N\langle E_{kin}\rangle$. Following \cite{fried04}, the quantity $\langle E_{kin}\rangle$ can be expressed as an expectation in terms of the probability distribution $ p(x) = \sum_n \psi_n^*\psi_n$.  Substituting in (\ref{eqn:epi}) and working with the single-dimension case for simplicity of analysis, we obtain
\begin{align*}
F' &= 4N\int{\sum_n\left|\frac{d\psi_n(x)}{dx}\right|^2~dx} + \lambda\left(\bar{E}_{kin} - C\int{E_{kin}\sum_n |\psi_n(x)|^2~dx}\right),\\
&= 4N\int{\sum_n\left|\frac{d\psi_n(x)}{dx}\right|^2~dx} - \lambda C\int{[W - V(x)]\sum_n |\psi_n(x)|^2~dx} + \lambda\bar{E}_{kin}.
\end{align*}
Applying the Euler-Lagrange equation $$\frac{d}{dx}~\frac{\partial \mathcal{L}}{\partial\psi'} = \frac{\partial\mathcal{L}}{\partial\psi},$$ with $\mathcal{L} = F'$, we get the solution,
\begin{equation}
\psi_n^{''}(x) + \frac{\lambda C}{4}[W - V(x)]\psi_n(x) = 0, \quad n = 1,\cdots N/2.
\label{eqn:schrodinger}
\end{equation}
Setting $\lambda C = \frac{8m}{\hbar^2}$ recovers a physically meaningful Schrodinger's equation without time dependence.

The general EPI approach has been used in several physical, biological and economics applications over the past decade with varying degrees of success. However, it has not acquired mainstream acceptance as a consequence of the arbitrary formulaic manner in which it has been presented and the \textit{ad hoc} metaphysical assumptions required to phrase it. It is hoped that the derivation of the formal structure of EPI with our sparse set of assumptions might occasion a re-evaluation of its utility.

\section {Discussion}
\label{sec:discuss}
Here, we summarize and discuss all the assumptions that we make in the process of deriving physical laws from the assumption of the ontological primacy of information (subjectively, awareness).

\begin{itemize}
\item{\textbf{Complete system description:} In the first instance, deriving physical laws appears to require that $s\in\S \Rightarrow s\in S'$, since the probability assignments $p[s]$ to outcomes $s$ and all subsequent calculations are likely to be flawed irretrievably otherwise. This distinguishes physical laws from general statistical inference where $\S'\subset\S$. It must be clarified here that the use of the term \emph{complete} to refer to the specification of set $\S$ is a slight misnomer, since we appear to be attaching ontological reality to $\S$ independent of the observer. This is partially true. While we could redress this concern by allowing a superset of $\S$ to replace $\S$ (with $\S$ becoming yet another $\S'$ in the process) whenever new information about the space of outcomes arises, such an approach allows for an unintuitive infinite regress. It is felt that it is simpler to allow $\S$ to have some ontological reality by considering it to be the space of outcomes observable to a universal observer with (or holding the potential for)\emph{complete} knowledge. The ontological realism of this observer itself is a separate (and philosophically important) question, but one that need not be addressed for the mathematical consistency of our framework.}

\item{\textbf{Bayesian inference:} We have assumed that observers update their beliefs about the possibility of various outcomes occurring by observing the system's behavior over multiple instances. While any other system of inference would be equally acceptable, for our demonstration of the emergence of physical laws, we have selected Bayesian inference as the specific model. It would be very surprising, in the author's opinion, were another consistent system of inference to result in results radically different from the Bayesian approach. Thus, we feel that it is sufficiently general to require no additional justification.}

\item{\textbf{Agnosticism about outcomes:}} Since we have worked throughout our exposition with either the MaxEnt specialization of Bayesian inference or an equivalent Fisher Information minimization, we have assumed that observers do not assume any knowledge unjustified by observations.

\item{\textbf{Existence assumption:} In the case where individual probabilities can be assigned to system outcomes in a physically interpretable manner (Section \ref{sec:shannon}) we have assumed that observers constructing physical laws believe they know of all possible outcomes that constitute the domain of the system.}

\item{\textbf{Measurements are accurate on average:} In all cases, we have assumed that observers believe they have estimated the behavior of the system accurately through a series of measurements upon various outcomes of the system.}

\item{\textbf{Physical constants:} Where appropriate, we have substituted physically meaningful quantities for scaling parameters and variables (e.g. Boltzmann's constant, Planck's constant). We do not claim any \textit{a priori} reason for these substitutions, but suggest that their introduction in non-informational physics derivations tend to be no less arbitrary. Unfortunately, the informational physics framework does not appear to be any closer than standard formulations in solving the \emph{fine-tuning}\footnote{Why do the universal physical constants have the values they do?} problem.}
\end{itemize}

Subjectively speaking, this set of assumptions appears to be quite reasonable, sparse enough to allow the informational physics principle to be enunciated, yet not so sparse as to require suspension of disbelief in order to accept the principle's implications. In plain language, the informational physics principle, in view of these assumptions, is suggesting the following: take a set of observers with complete knowledge of the range of behavior of a phenomenon under observation and give them sufficient observations to estimate some quantity that summarizes the past behavior of all outcomes of the system. While trying to infer a concise statement that would allow them to (a) summarize their findings and (b) make predictions about the system's behavior, they will all find the same mathematical structures as statements of physical laws, a circumstance that emerges from the nature of the process of observation and no ontological reality apart from the observer's perception.

\subsection{Future prospects}
\label{sec:prospects}
Much of the circumspection that arises with respect to informational physics derives from the insubstantial and intangible nature of its constituent categories. That is, it is far simpler to think in terms of electrons than bits, since one can measure the former's effects in the physical realm. Almost definitionally, the material (dualistic) definition of physics is far more amenable to reductionist manipulation than the informational (non-dualistic) definition. It is almost certain that this will remain a critical bottleneck to the emergence of informational physics as a mainstream discipline. However, we contend that demonstrating the consistency and duality of informational physics with respect to traditional physics is still not an enterprise without value. We examine and motivate future efforts along this line of inquiry in the following three contexts:

\begin{itemize}
\item {\textbf{Practical implications:}} From the practical point of view, the informational view of physics argues for a reappraisal of developments in approximation techniques in the light of the existence and continual development of statistical learning algorithms for different domain representations in the machine learning community. At the very least, the inference method can guide selection of the appropriate class of analytical methods appropriate for a particular problem domain. The first component of using the informational framework in physically realistic setting is \emph{domain specification}, which is equivalent to specifying an interesting and non-trivial domain of outcomes $\S$ of corresponding to physical measurements (for either single particle or ensemble settings). The critical element in performing inference, once the domain $\S$ has been specified, is the imposition of an appropriate measure to generate meaningful $p[s]$ corresponding to $s\in\S$, which should be a problem of some interest to measure theoreticians. The rest of the analysis can proceed using well known methods in both the statistics and computational learning literature.

\item{\textbf{Theoretical implications:}} Much has been made of the necessity of reaching a deeper understanding of the seemingly counter-intuitive foundations of quantum mechanics. Furthermore, in recent years, there has been a backlash in some segments of the theoretical physics community against the predominance of mathematical manipulation detached from reality. Informational physics carries the potential of remedying both these problems, as well as scope for a unification framework potentially more aesthetically satisfying than any action string theory might hope to derive. In the case of quantum mechanics, the irreconcilability of local realism with quantum predictions is trivially resolved by allowing quantum histories to emerge subjectively (an interpretation which is compatible with relational quantum mechanics \cite{rove96}). Furthermore, observer-dependent wave function collapse can now be interpreted as an update of outcome probabilities based on information acquisition. The mathematics of the informational framework of physics proceeds from well-understood and motivated assumptions regarding the nature of reality, which places it favorably in contrast to some more mathematically sophisticated but less simply interpretable systems.    

\item{\textbf{Epistemological implications:}} As our final point, we ask the following related questions: what are the implications of an informational basis of reality as well as knowledge about reality (the laws of physics being a subset consisting of statements regarding ideal scenarios) on (a) the prospect of unification, (b) the prospect of understanding the nature of physical laws and (c) the prospect of understanding the nature of the comprehensibility and predictability of reality?

Considering these questions one by one, it is not evident that the informational framework offers much greater hope for unification in the traditional sense of the term. While the informational framework is extremely general, the interpretability of its conclusions in the physical domain cannot occur arbitrarily. Should advances in \emph{domain specification} be feasible, the mathematical power of the informational framework will exceed that of the standard system. Until then, unification will continue to be equally inaccessible to either paradigm, though somewhat less philosophically necessary for the former than the latter.

Regarding the prospect of understanding the nature of physical laws, the informational framework claims a nearly complete resolution of this problem by showing that they emerge as special cases of a general inference procedure. The substitution of variables with specific values however, leaves it open to the fine-tuning criticism, as addressed earlier. The substitution of variables with quantities that appear to have specific physical interpretations leaves the framework open to the criticism of non-falsifiability. Simply put, the informational argument can retrieve existing laws and potentially find epistemologically-motivated approximations for them, but it cannot find completely new laws, since the variables that emerge would have no defined physical interpretation. Since it becomes impossible to construct a scenario where the informational framework constructs a law that is `false', the problem of non-falsifiability arises. This is a strong argument, and one that we are not in a position to refute at the moment. It should, therefore, be borne in mind while considering the plausibility of the informational physics hypothesis.

To answer our final question, we consider the promise that informational physics holds in answering Wigner's question \cite{wign60}. An assumption that reality is foundationally based in information (subjectively, awareness) dovetails quite well with the evidence we have presented in favor of the hypothesis that knowledge-acquisition about system outcomes is the foundation of physical laws. Since the informational physics principle declaims against the ontological reality of observer-independent systems, it explains Wigner's observation regarding the unreasonable effectiveness of mathematics as a descriptor of physical reality in the following way: observers construct laws of physics based on their information extremization apparatus (calculus of variations etc.). All observers who make the assumptions described in Section \ref{sec:discuss} will discover the same mathematical (and cognitive) statements as physical laws. This will make the relevant mathematics appear to be unreasonably effective in describing what all the observers \emph{believe-they-know}\footnote{Drawing a parallel to Plato's Theory of Forms is irresistible!}.
\end{itemize}

Thus, notwithstanding the occasional ambiguity that must necessarily ensue in treating with slippery concepts like information and awareness while attempting to describe reality, there appear to be strong physical, mathematical and epistemological reasons for further investigating the informational physics hypothesis. Establishing the validity of this hypothesis is a necessary step in facilitating a meaningful discussion of the broader consideration of awareness being the primal ontological entity.















\end{document}